\documentclass[fleqn,twoside]{article}
\usepackage{espcrc2}

\newcommand{\Dslash}{\slash \hspace{-7pt} D}
\newcommand{\dslash}{\slash \hspace{-5pt} \partial}
\newcommand{\pslash}{\slash \hspace{-5pt} p}
\newcommand{\cngi}{c_{\rm NGI}}
\def\Tr{\mathop{\mbox{Tr}}}

\def\tilde{\widetilde}

\def\){\right)} 
\def\({\left(} 
\def\]{\right]} 
\def\[{\left[}

\newcommand{\refeq}[1]{(\ref{eq:#1})}

\newcommand{\Eq}{Eq.~\refeq} 

\newcommand{\beq}{\begin{eqnarray}}
\newcommand{\eeq}{\end{eqnarray}}

\begin{document}

\title{
Improved non-perturbative renormalization without $\cngi$
}

\author{Stephen R. Sharpe
\address{Physics Department, Box 351560,
University of Washington, Seattle, WA 98195-1560, USA}%
\thanks{Email: sharpe@phys.washington.edu. 
Supported by DOE contract DE-FG03-96ER40956/A006.} 
}
      
\begin{abstract}
Recently, 
a method for $O(a)$ improvement of composite operators has been
proposed which uses the large momentum behavior of fixed gauge quark
and gluon correlation functions (G. Martinelli {\em et al.},
hep-lat/0106003). A practical problem with this method is that
a particular improvement coefficient, $\cngi$, which has a
gauge non-covariant form, is difficult to determine.
Here I work out the size of the errors made in improvement
coefficients and physical quantities if one does not include
the $\cngi$ term.
\end{abstract}

\maketitle

Non-perturbative renormalization~\cite{NPR} (NPR) is a method for
determining the renormalization constants for composite operators
which does not rely on perturbation theory.
In a recent paper, the method was extended to allow
the determination of the improvement coefficients needed to
remove all $O(a)$ corrections from matrix elements of bilinear 
operators~\cite{impNPR}. This improvement to NPR is important because
it allows the determination of fully $O(a)$ improved and renormalized
composite operators even for operators with non-vanishing anomalous
dimensions, e.g. scalar and pseudoscalar bilinears.

The NPR method uses amputated quark and gluon correlation functions,
determined in a fixed gauge, usually Landau gauge.
To improve the method one must first improve the quark
field itself, which, it is argued in Ref.~\cite{impNPR}, 
requires\footnote{%
Here $am=1/2\kappa-1/2\kappa_c$.
This and subsequent equations are valid only up to corrections
of $O(a^2)$.}
\beq
\hat q = \left(Z_q^0\right)^{-1/2}\left(1 \! -\! \frac{a}{2} b_q m 
\!+\! a c'_q W\! +\! a \cngi\, \dslash \right) q\,.
\eeq
Here $W\!=\!\Dslash+m_0$ is the entire $O(a)$ improved Wilson action
(including the clover/SW term). There are thus three independent
improvement coefficients, $b_q=1 + O(\alpha_s)$,
$c'_q=-0.25+O(\alpha_s)$ and $\cngi=O(\alpha_s)$. The latter
multiplies a gauge non-covariant term, but is nevertheless consistent
with BRST symmetry. It turns out to vanish at tree level, but its
one-loop value is not known---the one-loop calculation of the
propagator in Ref.~\cite{Capitani} only determines two linear
combinations of the three coefficients.

While, in principle, the three improvement coefficients can be
determine separately, this is difficult in practice~\cite{Becirevic}.
To understand this, note that the improved quark propagator is given
by~\cite{impNPR}
\beq
Z_q^0 \hat S = S_L\! +\!
 2 a c'_q 
\!+\! 2 a \cngi S_L i \pslash \!-\! a b_q m S_L,
\label{eq:impS}
\eeq
where $S_L$ is the bare lattice propagator.
To determine $\cngi$ and $c'_q$ one enforces, for large $p$,
\beq
\Tr \hat S(p) = 0 \,,
\label{eq:TrS}
\eeq
since, in the continuum, chiral symmetry
ensures that the r.h.s. of (\ref{eq:TrS})
is asymptotically proportional to $m/p^2$ (up to logarithms).
As can be seen from \Eq{impS}, the $c'_q$ term gives a constant
($p$-independent) contribution to $\Tr \hat S$, which is to be
chosen to cancel a corresponding contribution from $S_L$.
To evaluate the $\cngi$ term, we can replace $S_L$ by
$Z_q^0 \hat S$ (since $O(a^2)$ terms are not controlled),
and use the fact that, for large $p$,
$\hat S= 1/[i\pslash \Sigma_1(p^2)]$.
Thus the $\cngi$ term is proportional to $a/\Sigma_1(p^2)$.
This is not a constant, and thus differs, in principle,
from the $c'_q$ term.
The two contributions are difficult to separate in practice, however,
since $\Sigma_1(p^2)$ only varies logarithmically with $p$.

Were one able to determine $c'_q$ and $\cngi$, the remaining
improvement coefficient $b_q$ and the normalization $Z_q^0$ could
be determined by enforcing
\beq
\Tr \gamma_\nu \hat S(p)^{-1}\bigg|_{p^2=\mu^2} = i \Tr\gamma_\nu\ 
\pslash \,,
\label{eq:Scond2}
\eeq
where $\mu$ is a renormalization scale chosen 
so that sub-leading terms proportional to powers 
of $m/\mu$ and $\Lambda_{\rm QCD}/\mu$ can be ignored.\footnote{%
For later algebraic convenience, I have used
a different convention for $Z_q^0$ than the standard choice
of Ref.~\cite{NPR}, which involves derivatives.}
If we decompose the inverse propagator as
\beq
\hat S(p)^{-1} = i \pslash \Sigma_1(p^2) + \Sigma_2(p^2) + O(a^2) \,,
\eeq
then $\Sigma_1(\mu^2)=1 + O(m^2/\mu^2)$. Note the presence of sub-leading
terms in $\Sigma_1$, 
which are physical, and will play an important role below.
A possible definition of the improved, renormalized quark mass
is then $m_R = \Sigma_2(\mu^2)$.
With these definitions, I note that
\beq
\hat S(p)^{-1}\bigg|_{p^2=\mu^2} = i \pslash \!+\! m_R \!+\! 
O(a^2,\pslash m^2/p^2)\,.
\label{eq:impSatmu}
\eeq

Now I return to the main question:
What errors are induced  if we set $\cngi=0$, 
but keep all other improvement terms?
Then we have an approximately improved field
\beq
\tilde q = \left(\tilde Z_q^0\right)^{-1/2}\!
\left(1 - \frac{a}{2} \tilde b_q m 
+ a \tilde c'_q (\Dslash+m_0) \right) q \,.
\eeq
I will denote the
differences between true and approximate coefficients as follows:
\beq
\Delta b_q = \tilde b_q - b_q\,,\quad
\Delta c'_q = \tilde c'_q - c_q\,,\ \  {etc.}
\eeq
The idea is then to adjust
the approximate improvement constants so as to compensate,
to the extent possible, for the absence of the $\cngi$ term. 
In particular, I propose enforcing
\beq
\Tr \tilde S(p) \bigg|_{p^2=\mu^2, m=0} = 0 \,,
\label{eq:newTrS}
\eeq
which is a weaker version of \Eq{TrS},
and
\beq
\Tr \gamma_\nu \tilde S(p)^{-1}\bigg|_{p^2=\mu^2} = i \Tr\gamma_\nu\
\pslash \,,
\label{eq:newScond2}
\eeq
which has the same form as \Eq{Scond2}.
Since both conditions are satisfied by $\hat S$,
they apply as well to $\Delta S=\tilde S- \hat S$.

It is straightforward to work out the implications of these new
conditions using the result
\beq
\Delta S 
\!=\! 2a \left( {\Delta c'_q \over Z_q^0} \!-\! 
 [\Delta \ln Z_q^0 \!+\! m \Delta b_q 
  \!+\! \cngi i \pslash] \hat S \right)\!,
\label{eq:DeltaS}
\eeq
with $\cngi$ being the correct value of this improvement coefficient.
Enforcing Eqs.~(\ref{eq:newTrS},\ref{eq:newScond2}) leads to\footnote{%
The first two equations were also obtained in Ref.~\cite{Becirevic}.}
\beq
\Delta c'_q \!=\! Z_q^0 \cngi\,,\
\Delta \ln Z_q^0 \!=\! 0 \,, \
\Delta b_q \!=\! 2 \cngi m_R/m \,.
\nonumber 
\eeq
Since $m_R/m \equiv Z_m(1 + a b_m m)$ is of $O(1)$ in perturbation
theory, as is $Z_q^0$, we see that the errors in  
$c'_q$ and $b_q$ are proportional to $\cngi$, and thus of
$O(\alpha_s)$. This turns out to be generic for ``unphysical''
improvement coefficients, i.e. those that do not enter into on-shell
matrix elements of gauge invariant operators.
By contrast, there is no error in $Z_q^0$: ignoring $\cngi$ leads to
the correct $O(a)$ improved normalization in the chiral limit.
This is also generic: bilinears are $O(a)$ improved 
in the chiral limit even if $\cngi$ is ignored.

Using the results for the errors in improvement coefficients, 
the error in the 
inverse propagator is
\beq
\Delta[S^{-1}] \!=\! 2 a \cngi \hat S^{-1} 
\left( i\pslash \!+\! m_R \!-\! \hat S^{-1} \right) \,.
\label{eq:DeltaSinvres}
\eeq
From this I conclude that, at generic large $p$, ignoring $\cngi$
leads to an $O(a)$ error in the propagator. This is as expected
since we are performing an approximate improvement.
However, for $p^2=\mu^2$, where $\hat S^{-1}=i\pslash + m_R$, it
appears that $\Delta S^{-1}= O(a^2)$. This would be very surprising,
since it would allow the extraction of the exact improved renormalized quark
mass---a physical quantity---in an approximate improvement scheme.

The resolution 
is that one must include sub-leading 
terms in $\hat S^{-1}$ when evaluating \Eq{DeltaSinvres}:
\beq
\hat S(p)^{-1}\bigg|_{p^2=\mu^2} = i\pslash \, [1 + O(m^2/p^2)] 
+ m_R \,.
\eeq
The subleading terms are physical, i.e. not lattice artifacts, 
but in perturbation theory
they are of $O(\alpha_s^2)$ in Landau gauge~\cite{Capitani}.
One then finds
\beq
\Delta[S(p)^{-1}] \propto  a \cngi m^2 O(\alpha_s^2)
\eeq
which leads to 
\beq
\Delta m_R \propto a \cngi m^2 O(\alpha_s^2) \Leftrightarrow
\Delta b_m \propto \cngi O(\alpha_s^2) \,. \nonumber
\eeq
Thus the physical improvement coefficient $b_m$ does have an error if
one ignores $\cngi$, but this is of $O(\alpha_s^3)$ and presumably
much smaller than those noted above in unphysical coefficients.

The only good news concerns $\Sigma_1(p^2)$. Since this is
independent of $m$ at large $p$, it can be evaluated from the
(correctly improved) propagator {\em in the chiral limit}. 
But in this limit one finds (from \Eq{DeltaSinvres}) 
\beq
\Delta [S(p,m=0)^{-1}] = 2 a \cngi p^2 [\Sigma_1(p^2)-1]\Sigma_1(p^2)\,.
\nonumber
\eeq
Thus there is no $O(a)$ error in $\tilde\Sigma_1$,
the $\pslash$ part of $\tilde S^{-1}$,
if it is evaluated in the chiral limit:
\beq
\tilde \Sigma_1(p^2) = \Sigma_1(p^2) +O(am) + O(a^2) \,.
\eeq

It is straightforward to extend the analysis to bilinears.
The off-shell $O(a)$ improvement of bilinears involves only gauge
invariant operators, so there are no other improvement coefficients
analogous to $\cngi$. In particular, one can work through the
procedure for improving bilinears using amputated correlation
functions laid out in Ref.~\cite{impNPR}, but using the approximately
improved propagator $\tilde S$ instead of $\hat S$. I do not have
space to present the details here, and report only the conclusions.
As above, one must be careful to keep sub-leading contributions
to physical vertices, or one can end up fooling oneself that 
$O(a)$ errors can be avoided at special kinematic points.

For the scalar bilinear, off-shell improvement requires using
\beq
\hat S(x) = Z_S^0 (1 + a b_S m) [\bar q q(x) + a c'_S E_S(x)]\,,
\eeq
where $E_S$ is an operator, defined in Ref.~\cite{impNPR},
which vanishes by the lattice equations of motion, and thus leads
only to contact terms. On-shell improvement requires knowledge
only of $b_S$, and the normalization constant $Z_S^0$.
I find that the error in these constants made by ignoring $\cngi$ is
\beq
(\Delta c'_q \!+\! \Delta c'_S) Z_S^0 \!=\! \cngi\,,\  \
\Delta \ln Z_S^0\! =\! 0\,,\nonumber \\
{\rm and}\ \  \Delta b_S \!\propto\! \cngi O(\alpha_S^2)\,.
\eeq
Combining these with the previous result for $\Delta c'_q$, 
I find the same heirarchy of errors
as deduced from the propagator analysis. The error in the
off-shell improvement coefficient $c'_S$ is ``large'', of
$O(\alpha_s)$, while that in the on-shell coefficient $b_S$ is
``small'', of $O(\alpha_s^3)$, and that in the normalization vanishes.
That $\Delta b_S$ is proportional to $\alpha_S^3$ and not $\alpha_s^2$ is
due to the fact that the sub-leading $m \pslash/p^2$ corrections to
the scalar vertex vanish at one-loop order in Landau gauge.

For all the other bilinears, the analysis is more complicated, because
the determination of all the improvement coefficients requires
applying one improvement condition at non-forward momenta. This
introduces the dependence of vertices on momentum transfer,
and, it turns out, leads to the generic expectation that
$\Delta b_\Gamma \sim \Delta c_\Gamma \propto \cngi O(\alpha_s)$,
one power of $\alpha_s$ larger than above. One also finds that
$\Delta \ln Z_\Gamma^0=0$.

In summary, ignoring $\cngi$ leads to errors in all improvement
coefficients, as expected,
but these errors are quite small for those needed for calculations
of physical matrix elements. It may be that other sources of
systematic error, such as the need to subtract large $O(a^2 p^2)$
corrections, exceed the error made by ignoring $\cngi$.
Furthermore, leaving out the $\cngi$ term has no effect on the
normalization constants of the operators in the chiral limit.

On the other hand, it is preferable to have a method without
uncontrolled errors. How can this be acheived using NPR, given
the difficulty in determining $\cngi$?
One method is to input knowledge of one or more improvement
coefficients obtained by other methods, e.g. that based on Ward
identities. For example, if one knew $b_m$, one could adjust $\cngi$
until the result for $m_R$ had the correct ratio of quadratic to
linear dependence on the quark mass $m$.
This seems to be equivalent to the method suggested here by
Bhattacharya~\cite{tanmoy}.

\vspace{-.1truein}
\section*{Acknowledgements}
\vspace{-.1truein}

I thank Roger Horsley, 
Giancarlo Rossi and Massimo Testa for comments and discussions.

\end{document}